\documentstyle[prl,twocolumn,aps]{revtex}
\begin{document}

\title{Myopic Bosonization}

\author{Girish S. Setlur}
\address{ Department of Physics and Materials Research Laboratory,\\
 University of Illinois at Urbana-Champaign , Urbana Il 61801}
\maketitle

\begin{abstract}
 As the title suggests, this is an attempt at bosonizing fermions in any number
 of dimensions without paying attention to the fact that the Fermi surface
 is an extended object. One is tempted to introduce
 the density fluctuation and its conjugate and recast the interacting
 problem in terms of these canonical Bose fields. However, we find that
 the attempt is short-sighted figuratively as well for the same reason.
 But surprisingly, this flaw, which manifests itself as an
 inconsistency between Menikoff-Sharp's construction of the kinetic energy
 operator in terms of currents and densities, and our ansatz for this operator,
 is nevertheless able to reproduce(although reluctantly)
 many salient features of the free theory.
 Buoyed by this success, we solve the interacting problem
 and compute the full propagator.
\end{abstract}

\section{Density Fluctuation and its Conjugate}

We showed earlier\cite{Setlur} that the field operator, may be expressed
in terms of the density fluctuation and its conjugate. 
\begin{equation}
\psi({\bf{x}}) = e^{-i\sum_{ {\bf{q}} }e^{i{\bf{q.x}}}X_{ {\bf{q}} }}
e^{i\sum_{ {\bf{q}} }e^{-i{\bf{q.x}}}
{\tilde{U}}_{0}({\bf{q}})\rho_{ {\bf{q}} }/N}\sqrt{\rho_{0}}
\end{equation}
where,
 $ [X_{ {\bf{q}} },\rho_{ {\bf{q}}^{'} }] = i\delta_{ {\bf{q}}, {\bf{q}}^{'} } $
 and all other commutators are zero.
 Actually, this concept of a conjugate to the density fluctuation 
 has appeared in many guises in the literature. Apparently, the first
 person to think of this in the context of Fermi systems was 
  Pines\cite{Pines}.
   Also, 
\begin{equation}
{\tilde{U}}_{0}({\bf{q}}) =
 (\frac{ \theta(k_{f} - |{\bf{q}}|) - w_{1}({\bf{q}}) }{ w_{2}({\bf{q}}) })
^{\frac{1}{2}}
\end{equation}
\begin{equation}
w_{1}({\bf{q}}) = (\frac{1}{4\mbox{ }N\mbox{ }\epsilon^{2}_{ {\bf{q}} }})
\sum_{ {\bf{k}} }(\frac{ {\bf{k.q}} }{m})^{2}
(\Lambda_{ {\bf{k}} }(-{\bf{q}}))^{2}
\end{equation}
\begin{equation}
w_{2}({\bf{q}}) = (\frac{1}{N})\sum_{ {\bf{k}} }
(\Lambda_{ {\bf{k}} }(-{\bf{q}}))^{2}
\end{equation}
here, $ \Lambda_{ {\bf{k}} }({\bf{q}}) = \sqrt{ {\bar{n}}_{ {\bf{k}}+{\bf{q}}/2 }(1 - {\bar{n}}_{ {\bf{k}}-{\bf{q}}/2 })} $ and
$ {\bar{n}}_{ {\bf{k}} } = \theta(k_{F}-|{\bf{k}}|) $.
Let us now postulate that the kinetic energy operator has the following form,
\begin{equation}
K = \sum_{ {\bf{k}} }\omega_{0}({\bf{k}})b^{\dagger}_{ {\bf{k}} }b_{ {\bf{k}} }
\label{KIN}
\end{equation}
$ \omega_{0}({\bf{k}}) = \epsilon_{ {\bf{k}} }/S({\bf{k}}) $ is the
 Bijl-Feynman dispersion\cite{Mahan}.
 Here $ \epsilon_{ {\bf{k}} } = k^{2}/2m  $ and $ S({\bf{q}}) = q/2k_{F} $
 is the static structure factor for small $ q $.
 The fields $ b_{ {\bf{k}} } $
 are canonical bosons. From this we may postulate,
\begin{equation}
X_{ {\bf{q}} } = \frac{i}{2\sqrt{N\mbox{   }S_{ {\bf{q}} }}
cos \theta_{ {\bf{q}} }}(b_{-{\bf{q}} }e^{i\theta_{ {\bf{q}} }}
 - b^{\dagger}_{{\bf{q}} }e^{-i\theta_{ {\bf{q}} }})
\end{equation}
also,
\begin{equation}
\rho_{ {\bf{q}} } = \sqrt{N\mbox{   }S_{ {\bf{q}} }}
(b_{ {\bf{q}} } + b^{\dagger}_{ -{\bf{q}} })
\end{equation}
here $ \theta_{ {\bf{q}} } $ is a nontrivial phase. These cannot be 
absorbed into the $ b_{ {\bf{q}} } $'s, if they were they would pop
up in the density fluctuation again.
The current density fluctuation has a similar form,
\begin{equation}
{\bf{j}}_{ {\bf{q}} } = {\bf{q}}\sqrt{\frac{N}{4S_{ {\bf{q}} }}}
(b_{ {\bf{q}} }-  b^{\dagger}_{ -{\bf{q}} })
\end{equation}
This form reproduces the commutator between the density-fluctuation and the
 kinetic energy,
\begin{equation}
[\rho_{ {\bf{q}} },K] = \sqrt{N\mbox{   }S_{ {\bf{q}} }}
\frac{ \epsilon_{ {\bf{q}} } }{S_{ {\bf{q}} }}
(b_{ {\bf{q}} } - b^{\dagger}_{ -{\bf{q}} })
 = {\bf{q}}.\frac{ {\bf{j}}_{ {\bf{q}} } }{m}
\end{equation} 
However, from the field operator, we also have,
\begin{equation}
{\bf{j}}_{ {\bf{q}} } = i{\bf{q}}N\mbox{     }X_{ -{\bf{q}} }
-i{\bf{q}}{\tilde{U}}_{0}({\bf{q}})\rho_{ {\bf{q}} }
\end{equation}
From these two we have,
\begin{equation}
tan \mbox{       }\theta_{ {\bf{q}} } = -2\mbox{    }
 {\tilde{U}}_{0}({\bf{q}}) \mbox{    }S_{ {\bf{q}} }
\end{equation}
\[
\psi({\bf{x}}) = e^{-\sum_{ {\bf{q}} }e^{i{\bf{q.x}}}
\frac{ b^{\dagger}_{ {\bf{q}} } }
{  2\sqrt{NS_{ {\bf{q}} }}  } }
 e^{ \sum_{ {\bf{q}} }e^{i{\bf{q.x}}}
\frac{ b_{ -{\bf{q}} } }{ 2\sqrt{NS_{ {\bf{q}} }} } }
\]
\begin{equation}
\times
e^{i\sum_{ {\bf{q}} } {\tilde{U}}_{0}({\bf{q}})/2N}
e^{-\sum_{ {\bf{q}} }\frac{1}{8NS_{ {\bf{q}} }} }\sqrt{\rho_{0}}
\end{equation}
The equal-time version of the propagator is,
\begin{equation}
\langle \psi^{\dagger}({\bf{x}})\psi({\bf{x}}^{'}) \rangle
 = \rho_{0}e^{\sum_{ {\bf{q}} }
(e^{i{\bf{q}}.({\bf{x}}-{\bf{x}}^{'})}-1)\frac{1}{4NS_{ {\bf{q}} }}}
\end{equation}
but we also know that,
\[
\langle \psi^{\dagger}({\bf{x}})\psi({\bf{x}}^{'}) \rangle
 = \frac{1}{V}\sum_{ {\bf{q}} }e^{i{\bf{q}}.({\bf{x}}-{\bf{x}}^{'})}
\theta(k_{f}-|{\bf{q}}|)
\]
\begin{equation}
 = \rho_{0} (1 +  \frac{1}{N}\sum_{ {\bf{q}} \neq 0 }
(e^{i{\bf{q}}.({\bf{x}}-{\bf{x}}^{'})}-1)
\theta(k_{f}-|{\bf{q}}|))
\end{equation}
Therefore,
\[
Ln(\langle \psi^{\dagger}({\bf{x}})\psi({\bf{x}}^{'}) \rangle)
\]
\begin{equation}
 \approx ln(\rho_{0}) +  \frac{1}{N}\sum_{ {\bf{q}} \neq 0 }
(e^{i{\bf{q}}.({\bf{x}}-{\bf{x}}^{'})}-1)
\theta(k_{f}-|{\bf{q}}|))
\end{equation}
Since
\begin{equation}
\frac{1}{4S_{ {\bf{q}} }} \neq \theta(k_{f}-|{\bf{q}}|))
\end{equation}
 the two propagators don't agree. But then we proceed since the title suggests
 that this is a short-sighted endeavour anyway.
 If we compute the full propagator
 and then multiply and divide by the free propagator and in
 the division use the form predicted by myopic bosonization and in the 
 numerator use the form predicted by elementary considerations then,
 even though all is not well, we should be still alive at the end of the
 day, albeit battered and bruised. Let us introduce an interaction
\begin{equation}
H_{I} = \sum_{ {\bf{q}} \neq 0 }\frac{v_{ {\bf{q}} } }{2V}
(NS_{ {\bf{q}} })(b_{ {\bf{q}} } + b^{\dagger}_{ -{\bf{q}} })
(b_{ -{\bf{q}} } + b^{\dagger}_{ {\bf{q}} })
\end{equation}
From this we may diagonalise the full problem in terms of new Bose
 fields $ d_{ {\bf{q}} } $,
\begin{equation}
H = \sum_{ {\bf{q}} }\omega_{ {\bf{q}} }d^{\dagger}_{ {\bf{q}} }d_{ {\bf{q}} }
\end{equation}
where,
\begin{equation}
\omega^{2}( {\bf{q}} ) = \omega^{2}_{0}( {\bf{q}} ) 
+ 2\rho_{0}v_{ {\bf{q}} }\omega_{0}( {\bf{q}} )
\end{equation}
and,
\begin{equation}
d_{ {\bf{q}} } = [d_{ {\bf{q}} },b^{\dagger}_{ {\bf{q}} }]b_{ {\bf{q}} }
 -  [d_{ {\bf{q}} },b_{ -{\bf{q}} }]b^{\dagger}_{ -{\bf{q}} }
\end{equation}
and,
\begin{equation}
b_{ {\bf{q}} } = [b_{ {\bf{q}} },d^{\dagger}_{ {\bf{q}} }]d_{ {\bf{q}} }
 -  [b_{ {\bf{q}} },d_{ -{\bf{q}} }]d^{\dagger}_{ -{\bf{q}} }
\end{equation}
\begin{equation}
 [d_{ {\bf{k}} },b^{\dagger}_{ {\bf{k}} }] =  \frac{\sqrt{NS_{ {\bf{k}} }}}
{V}v_{ {\bf{k}} }\frac{ [d_{ {\bf{k}} },\rho_{-{\bf{k}}}] }
{\omega({\bf{k}}) - \omega_{0}({\bf{k}})}
\end{equation}
\begin{equation}
 [d_{ {\bf{k}} },b_{ -{\bf{k}} }] = 
 -\frac{\sqrt{NS_{ {\bf{k}} }}}
{V}v_{ {\bf{k}} }\frac{ [d_{ {\bf{k}} },\rho_{-{\bf{k}}}] }
{\omega({\bf{k}}) + \omega_{0}({\bf{k}})}
\end{equation}
Since $ [d_{ {\bf{k}} },d^{\dagger}_{ {\bf{k}} }] = 1 $,
\begin{equation}
[d_{ {\bf{k}} },\rho_{-{\bf{k}}}]
 = \sqrt{ \frac{N}{ S_{ {\bf{k}} } } }
\sqrt{ \frac{ \omega_{0}(k) }{ \omega({\bf{k}}) } }
\end{equation}
Let us write down formulas for the propagators,
\[
\frac{ \langle \psi^{\dagger}({\bf{x}}^{'}t^{'})\psi({\bf{x}}t) \rangle }
{ \langle \psi^{\dagger}({\bf{x}}^{'}t^{'})\psi({\bf{x}}t) \rangle_{0} }
 = e^{-\sum_{ {\bf{q}} }\frac{1}{ 2NS_{ {\bf{q}} } }
[b_{ {\bf{q}} },d^{\dagger}_{ {\bf{q}} }][b_{ -{\bf{q}} },d_{ {\bf{q}} }] }
\]
\[
\times
e^{-\sum_{ {\bf{q}} }\frac{1}{ 2NS_{ {\bf{q}} } }
([b_{ {\bf{q}} },d_{ -{\bf{q}} }])^{2} }
\]
\begin{equation}
\times
e^{\sum_{ {\bf{q}} }
\frac{ e^{ i{\bf{q}}.({\bf{x}}-{\bf{x}}^{'})} }{4NS_{ {\bf{q}} }}
\{ \frac{ \omega_{0}({\bf{q}}) }{ \omega({\bf{q}}) }
e^{i\omega({\bf{q}})(t-t^{'})}
 - e^{i\omega_{0}({\bf{q}})(t-t^{'})} \} }
\end{equation}
\[
\frac{ \langle \psi({\bf{x}}t)\psi^{\dagger}({\bf{x}}^{'}t^{'}) \rangle }
{ \langle \psi({\bf{x}}t)\psi^{\dagger}({\bf{x}}^{'}t^{'}) \rangle_{0} }
 = e^{-\sum_{ {\bf{q}} }\frac{1}{ 2NS_{ {\bf{q}} } }
[b_{ {\bf{q}} },d^{\dagger}_{ {\bf{q}} }][b_{ -{\bf{q}} },d_{ {\bf{q}} }] }
\]
\[
\times
e^{-\sum_{ {\bf{q}} }\frac{1}{ 2NS_{ {\bf{q}} } }
([b_{ {\bf{q}} },d_{ -{\bf{q}} }])^{2} }
\]
\begin{equation}
\times
e^{\sum_{ {\bf{q}} }
\frac{ e^{i{\bf{q}}.({\bf{x}}^{'}-{\bf{x}})} }{4NS_{ {\bf{q}} }}
\{ \frac{ \omega_{0}({\bf{q}}) }{ \omega({\bf{q}}) }
e^{i\omega({\bf{q}})(t^{'}-t)}
 - e^{i\omega_{0}({\bf{q}})(t^{'}-t)} \} }
\end{equation}
 As far as the dielectric function is concerned, it may be evaluated 
 using methods outlined in our earlier work\cite{Setlur}.
 The final answer is as follows,
\begin{equation}
\epsilon({\bf{q}},\omega) = \frac{1}{1 + v_{ {\bf{q}} }
\rho_{0}S_{ {\bf{q}} }(\omega_{ {\bf{q}} }/\omega^{0}_{ {\bf{q}} })
(\frac{ 2\omega_{ {\bf{q}} } }{\omega^{2} - \omega^{2}_{ {\bf{q}} } })}
\end{equation}
 Assuming that $ \omega $ has a small imaginary part we recover damping as
 well. However, the above dielectric function apart from having obvious zeros
 at $ \omega = \omega_{ {\bf{q}} } $ does not seem to resemble the RPA
 dielectric function. This points to the fact that our approach
 is probably flawed. The main reason why our approach is wrong is because of the
 following reason. The Menikoff-Sharp
 construction\cite{Sharp} of the kinetic energy operator may be written as,
\begin{equation}
K = \int \frac{ d{\bf{x}} }{2m} (\frac{ {\bf{J}}^{2} }{\rho} 
+ \frac{ (\nabla\rho)^{2} }{4\rho})
 + c-number
\end{equation}
In momentum space it looks as follows(after expanding in powers of
 density fluctuations),
\begin{equation}
K = \frac{ ({\bf{j}}_{ {\bf{0}} })^{2} }{2mN}
+
 \sum_{ {\bf{q}}\neq 0 }
\frac{ {\bf{j}}_{ {\bf{q}} }.{\bf{j}}_{ -{\bf{q}} } }{2mN}
 +  \sum_{ {\bf{q}}\neq 0 }\frac{ \epsilon_{ {\bf{q}} }}{4N}
\rho_{ {\bf{q}} }\rho_{ -{\bf{q}} } 
\end{equation}
The current $ {\bf{j}}_{ {\bf{0}} } $ is the total current operator
 $ {\bf{j}}_{ {\bf{0}} } = \sum_{ {\bf{k}} }{\bf{k}}c^{\dagger}_{ {\bf{k}} }c_{ {\bf{k}} } $.
This may be written as,
\begin{equation}
{\bf{j}}_{ {\bf{0}} } =
 -\sum_{ {\bf{q}} }(i{\bf{q}})\rho_{ {\bf{q}} }X_{ {\bf{q}} }
\end{equation}
 If one uses the formulas for the density fluctuation and its conjugate
 in terms of the Bosons $ b_{ {\bf{q}} } $ we find
 that the resulting kinetic energy operator is not the one written
 down in Eq.(~\ref{KIN}).
 The only purpose of this exercise is to point out
 to future scholars that this route is a dead end.

\end{document}